\documentclass[letterpaper,english,reprint, aps]{revtex4-1}
\usepackage[T1]{fontenc}
\usepackage[latin9]{inputenc}
\usepackage{color}
\usepackage{textcomp}
\usepackage{mathrsfs}
\usepackage{amsmath}
\usepackage{amssymb}
\usepackage{graphicx}
\usepackage{esint}

\makeatletter

\pdfpageheight\paperheight
\pdfpagewidth\paperwidth

\makeatother

\usepackage{babel}
\begin{document}
\preprint{APS/123-QED}
\title{Performance optimization of a finite-time quantum tricycle}
\author{Jingyi Chen}
\affiliation{Department of Physics, Xiamen University, Xiamen 361005, People\textquoteright s
Republic of China.}
\author{Shihao Xia}
\affiliation{Department of Physics, Xiamen University, Xiamen 361005, People\textquoteright s
Republic of China.}
\author{Jincan Chen}
\email{jcchen@xmu.edu.cn}

\affiliation{Department of Physics, Xiamen University, Xiamen 361005, People\textquoteright s
Republic of China.}
\author{Shanhe Su}
\email{sushanhe@xmu.edu.cn}

\affiliation{Department of Physics, Xiamen University, Xiamen 361005, People\textquoteright s
Republic of China.}
\author{}
\date{\today}
\begin{abstract}
We establish a finite-time external field-driven quantum tricycle
model. Within the framework of slow driving perturbation, the perturbation
expansion of heat in powers of time can be derived during the heat
exchange processes. Employing the method of Lagrange multiplier, we
optimize the cooling performance of the tricycle by considering the
cooling rate and the figure of merit, which is the product of the
coefficient of performance and cooling rate, as objective functions.
Our findings reveal the optimal operating region of the tricycle,
shedding light on its efficient performance.
\end{abstract}
\maketitle

\section{INTRODUCTION}

With the aim of promoting the advancement of application development,
there has been a notable upsurge of interest in three-heat-source
systems\citep{key-Kosloff2013,key-Chen1989JCM,key-Qi2024}. These
models have attained substantial prominence within the field of thermodynamics.
Under the assumption of quasi-static conditions, a universal upper
bound on the coefficient of performance (COP) of a three-heat-source
refrigerator was derived \citep{key-Huang}. However, the quasi-static
assumption implies that the cooling rate of the system is zero, as
the execution of a practical cycle would necessitate a finite amount
of time. The problem that heat engines need non-zero power output
has sparked the emergence and advancement of a branch of thermodynamics
referred to as finite time thermodynamics \citep{key-denBroeck,key-IzumidaEurophys2012,key-IzumidaNJP2015}.
The optimized models within the realm of finite-time thermodynamics
have demonstrated the capability to achieve high efficent energy conversion
while simultaneously preserving power output or cooling rate \citep{key-denBroeck,key-Curzon,key-Salamon,key-Sekimoto,key-Esposito2009,key-Esposito2010,key-Schmiedl,key-Tu2008,key-Izumida2008,key-Izumida2009,key-Izumida2010}.

In the domain of finite-time thermodynamics, the quantitative analysis
of irreversibility plays a crucial role in the optimization of models
\citep{key-Curzon,key-Pekola,key-Vinjanampathy,key-Binderbook}. Regarding
this matter, Chen et al. derived the optimal relationship between
the cooling rate and COP for an endoreversible three-heat-source refrigerator
\citep{key-Chen1989}. As proposed by Esposito et al., under the conditions
of low dissipation, the entropy generation of a finite-time Carnot
cycle is inversely proportional to its cycle duration \citep{key-Esposito2010}.
Drawing inspiration from Ref. \citep{key-Esposito2010}, numerous
researchers have investigated the COP of low dissipation refrigerators
at the maximum cooling rate\citep{key-Tomas2012,key-Tom=0000E1s,key-Tu2012,key-Gonzalez-Ayala,key-Ye,key-yePRE}.
In particular, Guo et al. introduced and examined the performance
of a novel combined low dissipation three-terminal refrigerator model,
which was accomplished by coupling a low dissipation heat engine with
a low dissipation refrigerator \citep{key-guoECM,key-GUOJPA,key-ECM}.
Based on the various proposed finite-time thermodynamic models, significant
attention has also been directed towards three-heat-source systems
in the domain of quantum mechanics. This includes the exploration
of quantum absorption refrigerators\citep{key-PRL108070604(2012),key-PRE97052145(2018),key-PRE98012117(2018),key-PRE105034112(2022)},
self-consistent refrigerators\citep{key-PRE87042131(2013)}, and endoreversible
quantum refrigerators\citep{key-PRE90062124(2014)}. Although the
assumption of low dissipation has garnered significant attention,
the origin of irreversible entropy generation in three-heat-source
quantum systems and its impact on energy conversion have not been
deeply investigated.

The theory of open quantum systems offers a powerful framework for
investigating quantum effects in the realm of finite-time quantum
thermodynamics. The Markovian master equation approach, which has
emerged as a significant achievement in the study of open quantum
systems\citep{key-Breuerbook,key-Rivasbook}, offers a description
of the temporal evolution of a system subjected to a weak interaction
with environment. Cavina et al. developed a perturbation theory for
the quantum master equation with slowly varying parameters by employing
the Markovian master equation approach \citep{key-Cavina}. By analyzing
finite-time heat exchange processes, a notable correlation was discovered
between the first-order correction of heat and irreversible entropy
generation. Taking inspiration from Ref. \citep{key-Cavina}, several
researchers have proposed a universal framework for optimizing the
control of slow-driving quantum Carnot engines \citep{key-L124110606(2020),key-E98062132(2018)}.These
studies have highlighted the significance of quantifying irreversible
entropy generation, often achieved through the measurement of thermodynamic
length \citep{key-Scandi,key-Abiuso,key-ScandiPHD}. Chen et al. introduced
the concept of the Drazin inverse of the Lindblad superoperator to
characterize excess dissipation in quasistatic thermodynamic processes
\citep{key-Chenjf}. Su et al. utilized the Lagrange multiplier method
to optimize the cooling performance in a finite-time external field-driven
quantum refrigeration cycle \citep{key-PRE107044118(2023)}. 

In this study, we propose firstly a finite-time quantum tricycle (FTQTC)
model and employ the slow driving perturbation theory to determine
the expression for the first-order irreversible corrections of heat.
Furthermore, we employ the Lagrange multiplier method to optimize
the cooling performance of the quantum tricycle model. The explicit
determination of the form of irreversible entropy generation allows
for a unified optimization criterion in the investigation of low-dissipation
quantum tricycle models.

The paper is structured as follows: Section II introduces the FTQTC
model and utilizes the slow driving perturbation theory to derive
the first-order irreversible corrections of heat. Section III applies
the Lagrange multiplier method to determine the optimal performance
of the FTQTC in term of the cooling rate. In Section IV, we investigate
the suitable range of optimal performance in the FTQTC. Lastly, Section
V provides concluding remarks.

\section{THE CONTROL PROTOCOL OF A FINITE-TIME QUANTUM TRICYCLE}

As shown schematically in Fig. 1(a), a FTQTC is established through
a six-step process, directly inspired by its classical analogue \citep{key-guoECM}.
The working substance is a two level system (TLS) with time-dependent
Hamiltonian $H\left(t\right)=\hbar\omega_{v}\left(t\right)\sigma_{z}/2$,
where $\omega_{v}\left(t\right)$ is the energy splitting at time
$t$, $\sigma_{z}$ is the Pauli matrix in $z$ direction, and $\hbar$
is Planck's constant. The control protocol comprises three heat exchange
steps during which the TLS is placed in thermal contact with reservoir
$v\left(v=h,c,\mathrm{or}\,p\right)$, and three diabatic steps are
characterized by sudden quench in the Hamiltonian. 

A weak coupling between the system and reservoir $v$ is considered.
The density operator $\rho(t)$ of the TLS evolves according to the
Markovian master equation, i,e., $\frac{d}{dt}\rho\left(t\right)=\mathcal{L}_{v}\left(t\right)\left[\rho\left(t\right)\right]$,
where the generator $\mathcal{L}_{v}\left(t\right)$ represents the
quantum Liouvillian superoperator. By considering that $d\omega/dt$
is finite but small enough and introducing the dimensionless time-rescaled
parameter $s=t/\tau_{v}\left(s\in\left[0,1\right]\right)$ for a given
reservoir $v$, the first order approximation of the solution of the
density operator is given by 
\begin{figure}
\includegraphics[scale=0.5]{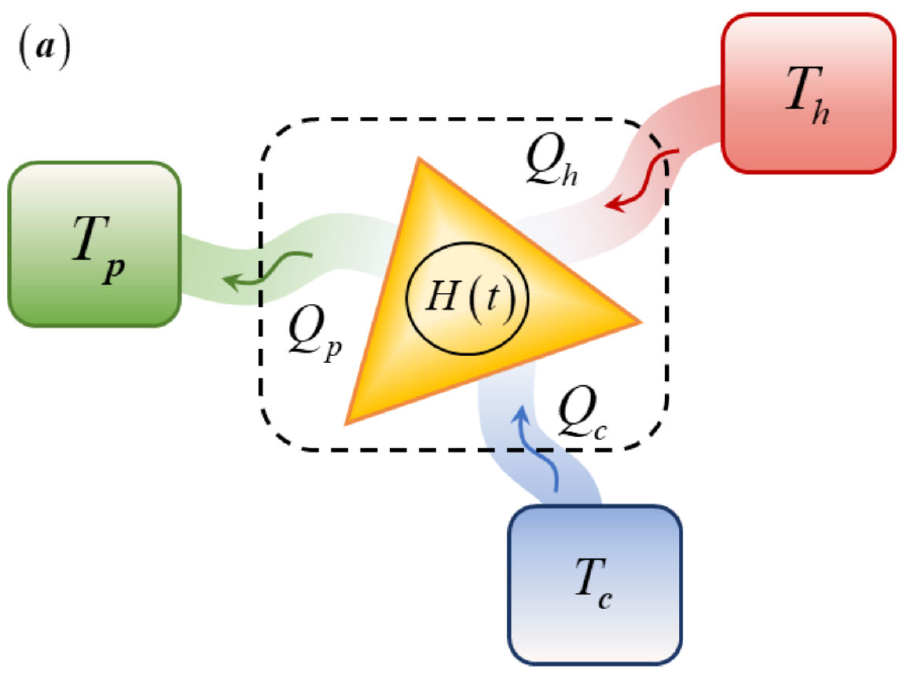}

\includegraphics[scale=0.3]{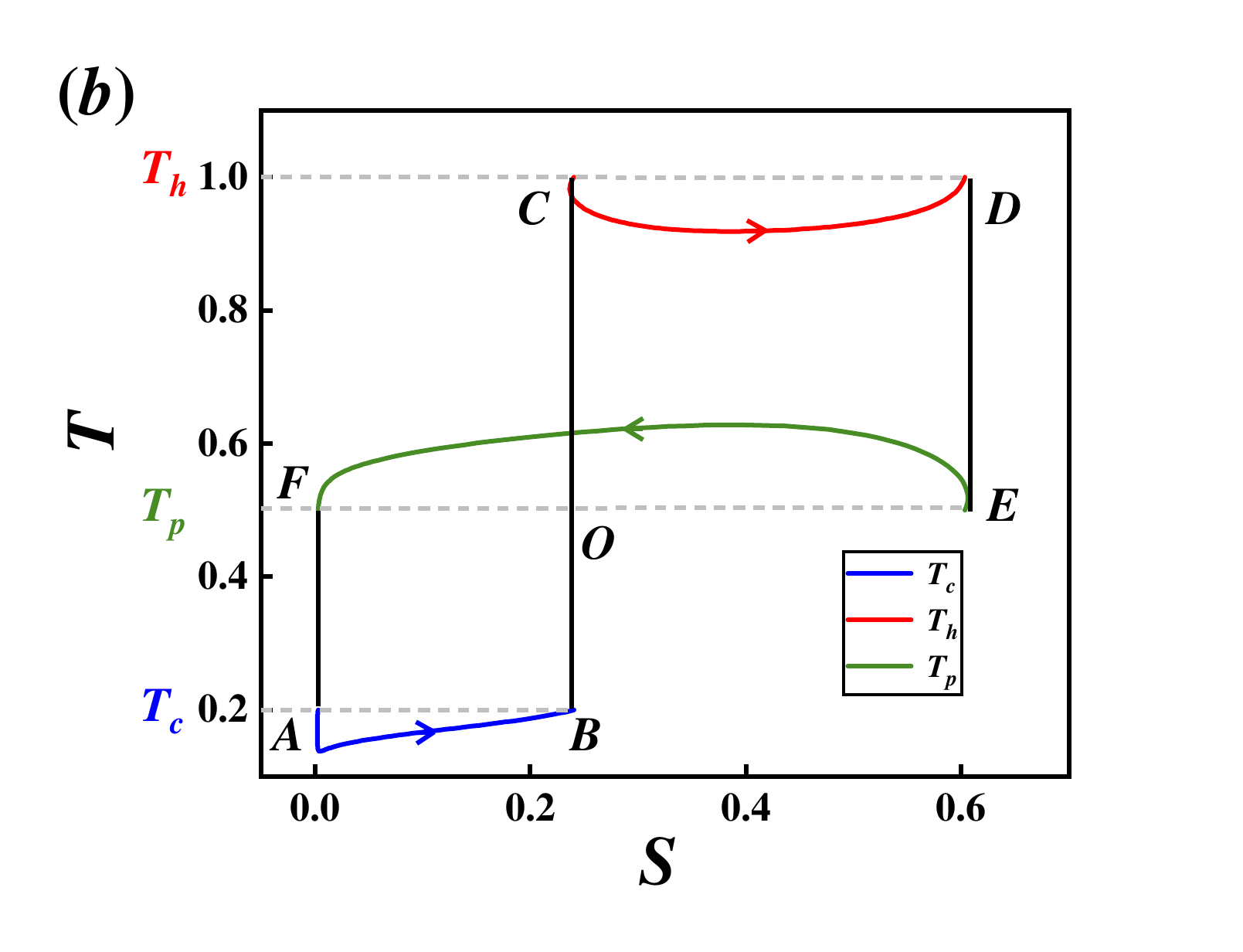}

\includegraphics[scale=0.3]{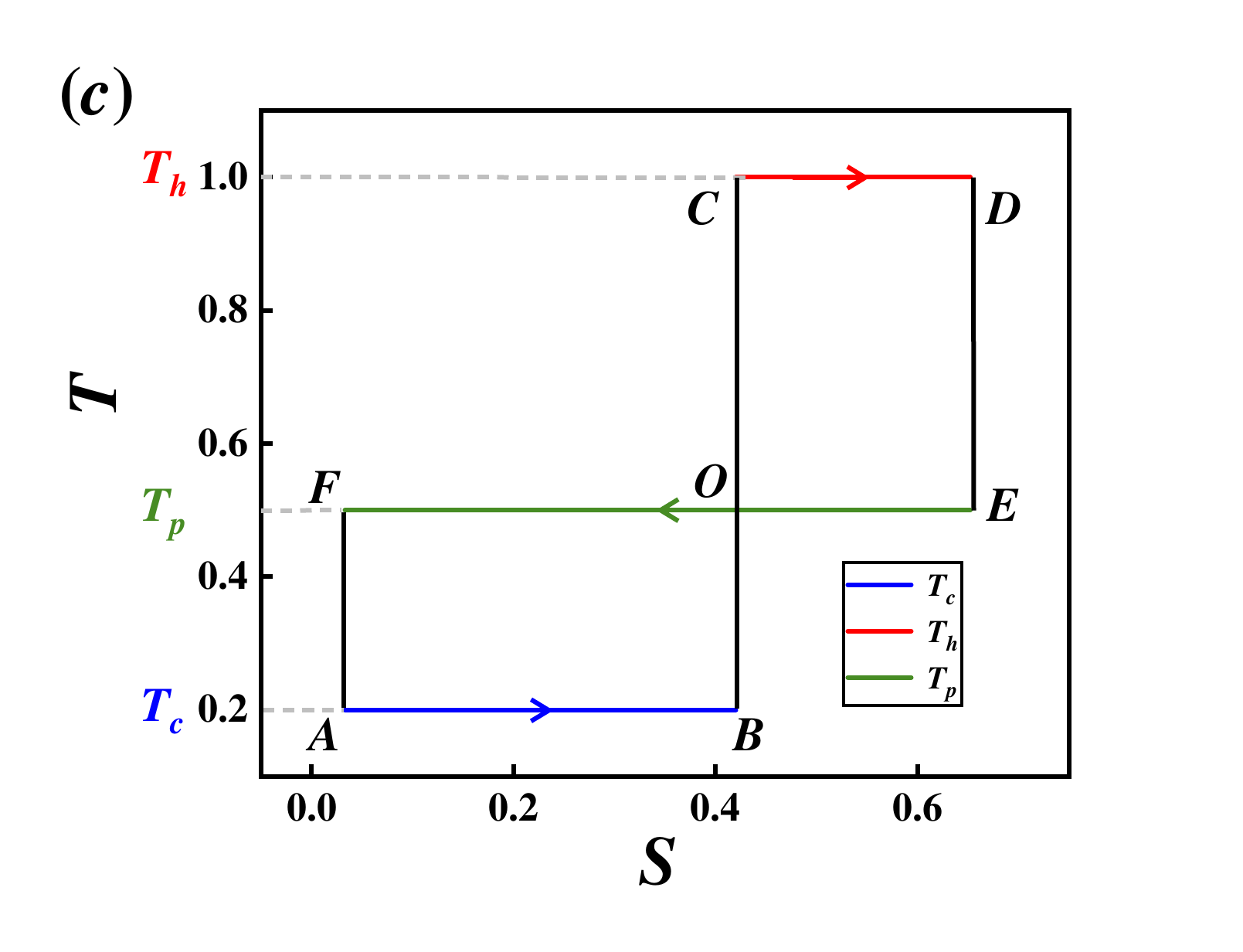}

\caption{(a) Schematic representation of a quantum tricycle. (b) The temperature-entropy
diagram of a FTQTC. (c) The temperature-entropy diagram of a reversible
quantum tricycle. $\delta_{c}=0.3492$ for the reversible model, while
$\delta_{c}=0.5333$ for the irreversible model. The values of the
remaining parameters: $T_{c}=0.2$, $T_{h}=1$, $T_{p}=0.5$, $\tau_{p}=11$,
$\tau_{c}=9$, $\alpha=0$, $\gamma_{0}=1$, and $\zeta_{c}=\zeta_{h}=2$.
The frequency exponent $\alpha$ is choose to be zero, which determines
the spectral density $J(\omega_{v}(t))\propto\left[\omega_{v}(t)\right]^{\alpha}$
of the bath. Unless otherwise specified, these parameters will be
used in the following figures. Planck\textquoteright{} s constant
$\hbar$ and Boltzmann\textquoteright{} s constant $k_{B}$ are set
to be unity throughout the paper.}
\end{figure}
\begin{equation}
\tilde{\rho}(s)=\tilde{\rho}_{\mathrm{eq},v}(s)+\frac{1}{\tau_{v}}\tilde{\mathscr{\mathcal{L}}}_{v}^{-1}\left(s\right)\frac{d}{ds}\left[\tilde{\rho}_{\mathrm{eq},v}(s)\right].\label{eq:q1-1}
\end{equation}
where $\tilde{\rho}(s)\equiv\rho(\tau_{v}s)$ , $\text{ }\tilde{\mathscr{\mathcal{L}}_{v}}^{-1}\left(s\right)\equiv\mathcal{L}_{v}^{-1}\left(\tau_{v}s\right)$
with $\mathscr{\mathcal{L}}_{v}^{-1}\left(\tau_{v}s\right)$ being
the Drazin inverse of $\mathcal{L}_{v}\left(t\right)$ \citep{key-Chenjf,key-Crook,key-Mandal},
and the instantaneous Gibbs state $\tilde{\rho}_{\mathrm{eq},v}(s)=\exp\left[-\tilde{H}\left(s\right)/\left(k_{B}T_{v}\right)\right]/\textrm{Tr}\left\{ \exp\left[-\tilde{H}\left(s\right)/\left(k_{B}T_{v}\right)\right]\right\} $
with $\tilde{H}\left(s\right)=H\left(\tau_{v}s\right)$. The details
of the calculation are reported in Appendix A. Based on Alicki\textquoteright{}
s definition of heat \citep{key-Alicki2016,key-Alicki2015} and Eq.(\ref{eq:q1-1}),
the amount of heat entering the system from reservoir $v$ during
the interval $\left[0,\tau_{v}\right]$ (Appendix B)
\begin{equation}
Q_{v}=Q_{v}^{0}+Q_{v}^{1}.\label{eq:Qc}
\end{equation}
The zeroth order approximation of the density operator recovers the
standard formula of heat in equilibrium thermodynamics, i.e., $Q_{v}^{0}=\beta_{v}^{-1}\Delta S_{\mathrm{eq},v}$,
where the entropy change of the equilibrium state of the system $\Delta S_{\mathrm{eq},v}=S_{\mathrm{eq},v}\left(\tau_{v}\right)-S_{\mathrm{eq},v}(0)$
with $S_{\mathrm{eq},v}(t)=-k_{B}\textrm{Tr}\left\{ \rho_{\mathrm{eq},v}(t)\ln\left[\rho_{\mathrm{eq},v}(t)\right]\right\} =-k_{B}\textrm{Tr}\left\{ \tilde{\rho}_{\mathrm{eq},v}(s)\ln\left[\tilde{\rho}_{\mathrm{eq},v}(s)\right]\right\} $.
The first order irreversible corrections of heat $Q_{v}^{1}=\beta_{v}^{-1}\Sigma_{v}/\tau_{\nu}$
with $\Sigma_{v}=\beta_{v}\int_{0}^{1}ds\textrm{Tr}\left[\tilde{H}\left(s\right)\frac{d}{ds}\left\{ \tilde{\mathscr{\mathcal{L}}}_{v}^{-1}\frac{d}{ds}\left[\tilde{\rho}_{\mathrm{eq},v}(s)\right]\right\} \right],$
and $\beta_{v}=1/\left(k_{B}T_{v}\right)$. For a TLS, the Liouvillian
superoperator $\mathcal{L}_{v}\left(t\right)$ , the instantaneous
equilibrium state $\rho_{\mathrm{eq},v}(t)$, and the Drazin inverse
$\mathscr{\mathcal{L}}_{v}^{-1}\left(t\right)$ of the dissipator
can be obtained from Appendix C, respectively. Figures 1(b) and (c)
show the temperature-entropy diagram of an irreversible FTQTC and
a reversible tricycle, respectively. The effective temperature $T_{\mathrm{eff},v}$
of the TLS can be calculated by using the energy eigenbasis. It is
given by the equation $T_{\mathrm{eff},v}=\frac{\hbar\omega_{v}\left(t\right)}{k_{B}}\left(\ln\frac{\rho_{0}}{\rho_{1}}\right){}^{-1}$,
where $\rho_{0}$ is the occupation probability of the ground state
and $\rho_{1}$ is the occupation probability of the excited state.
This equation is derived based on the assumption that the occupation
probabilities follow the Boltzmann distribution \citep{key-Quan}.
The entropy of a TLS can be calculated by using the density matrix
formalism, which is given by $S(t)=-k_{B}\textrm{Tr}\left\{ \rho(t)\ln\left[\rho(t)\right]\right\} =-k_{B}\textrm{Tr}\left\{ \tilde{\rho}(s)\ln\left[\tilde{\rho}(s)\right]\right\} $.
This entropy quantifies the amount of uncertainty or disorder in the
TLS and is based on the von Neumann entropy, which is a measure of
the system's mixedness or lack of pure state. The detail of the control
protocols of the quantum tricycle are designed as follows:

(1) Heat exchange with reservoir $c$ (from point A to B): When the
TLS is coupled to reservoir $c$ with low-temperature $T_{c}$, the
time-modulated field drives the frequency of the TLS according to
the cosine function $\omega_{c}\left(t\right)=\delta_{c}\left[\cos\pi\left(t/\tau_{c}\right)+\zeta_{c}\right]$
in the time interval $t\in\left[0,\tau_{c}\right]$, where $\delta_{c}$
and $\zeta_{c}$ denote the parameters of the amplitude and the displacement,
respectively. 

(2) Diabatic expansion (from point B to C): The system is isolated
from any reservoir and the frequency suddenly shifted from $\omega_{c}\left(\tau_{c}\right)$
to $\omega_{h}\left(0\right)=T_{h}/T_{c}\omega_{c}\left(\tau_{c}\right)$.
The scaling factor $T_{h}/T_{c}$ is introduced, ensuring that the
system continuously remains in equilibrium state during a reversible
quantum tricycle. In addition, the rapidly changing condition of diabatic
process prevents the TLS from adapting its configuration during the
process, and thus the probability density remains unchanged. 

(3) Heat exchange with reservoir $h$ (from point C to D): The system
is coupled to reservoir $h$ with high-temperature $T_{h}$ in the
time interval $t\in\left[0,\tau_{h}\right]$. The Hamiltonian $H\left(t\right)$
of the system is concurrently modified by adjusting the frequency
according to the function $\omega_{h}\left(t\right)=\delta_{h}\left[\cos\pi\left(t/\tau_{h}\right)+\zeta_{h}\right]$,
where $\delta_{h}$ and $\zeta_{h}$ signify the parameters of the
amplitude and the displacement, respectively. 

(4) First diabatic compression (from point D to E): The system is
decoupled from the reservoir and a sudden quench is performed, in
which the frequency is changed from $\omega_{h}\left(\tau_{h}\right)$
into $\omega_{p}\left(0\right)=\left(T_{p}/T_{h}\right)\omega_{h}\left(\tau_{h}\right)$. 

(5) Heat exchange with reservoir $p$ (from point E to F): An irreversible
heat exchange process follows by coupling the system with reservoir
$p$ at intermediate-temperature $T_{p}$. The frequency of the system
is slowly changed in accordance with the function $\omega_{p}(t)=\delta_{p}\left(\cos\pi\left(1-t/\tau_{p}\right)+\zeta_{p}\right)$
in a time interval $t\in\left[0,\tau_{p}\right]$. The parameters
$\delta_{p}$ and $\zeta_{p}$ are the amplitude and the displacement
in this process, respectively. 

(6) Second diabatic compression (from point F to A): Finally, the
system is disconnected from the reservoir and a sudden quench restores
the frequency back to $\omega_{c}\left(0\right)=\left(T_{c}/T_{p}\right)\omega_{p}\left(\tau_{p}\right)$. 

For a given heat exchange process, the time derivatives of $\omega_{v}\left(t\right)$
at the beginning and the end of time are equal to zero, i.e., $\dot{\omega}_{v}\left(0\right)=\dot{\omega}_{v}\left(\tau_{v}\right)=0$.
This also gives rise to $\frac{d}{ds}\left[\tilde{\rho}_{\mathrm{eq},v}(s)\right]=0$
at $s=0$ and $s=1$. Acccoring to Eq. (\ref{eq:q1-1}), it guarantees
that the system maintains the same equilibrium state both at the beginning
and at the end of the heat exchange process. Furthermore, it ensures
that the system remains in close proximity to the instantaneous steady
state throughout the entire cycle. The selection of the scale parameter
in frequency alteration during adiabatic operations enables a smooth
transition of the system between the instantaneous equilibrium states
associated with different reservoirs. With the given temperatures
of $T_{c}$, $T_{h}$, and $T_{p}$, as well as the values of $\zeta_{c}$,
$\zeta_{h}$, and $\delta_{c}$, we can deduce the following relationships:
$\zeta_{p}=\frac{1+\zeta_{c}\zeta_{h}}{\zeta_{c}+\zeta_{h}}$, $\delta_{h}=\frac{T_{h}\left(\zeta_{c}-1\right)}{T_{c}\left(1+\zeta_{h}\right)}\delta_{c}$,
and $\delta_{p}=\frac{T_{p}\left(\zeta_{c}+\zeta_{h}\right)}{T_{c}\left(1+\zeta_{h}\right)}\delta_{c}$
(Appendix D).

Based on the the temperature-entropy diagram shown in Figs.1(b) and
(c), it is observed that $Q_{h}^{0}>0$ , $Q_{c}^{0}>0$, and $Q_{p}^{0}<0$,
as $\Delta S_{\mathrm{eq},h}>0$, $\Delta S_{\mathrm{eq},c}>0$, and
$\Delta S_{\mathrm{eq},p}<0$. In Ref. \citep{key-PAbiuso}, it is
demonstrated that the first-order irreversible corrections of heat,
i.e., $Q_{v}^{1}\leq0$. The sum of the first-order irreversible corrections
of heat, represented by the expression $\sum_{v}Q_{v}^{1}$, is less
than or equal to zero. In addition, according to the principles of
energy conversion, the sum of the heat entering a system in a cycle
\begin{equation}
\sum_{v}Q_{v}=\sum_{v}Q_{v}^{0}+\sum_{v}Q_{v}^{1}=0.\label{eq:eb}
\end{equation}
Thus, the sum of the zeroth order approximation of heat $\sum_{v}Q_{v}^{0}\geq0$.
In a finite time cycle, it is true that $\sum_{v}Q_{v}^{1}\neq0$,
which implies that $\sum_{v}Q_{v}^{0}=\sum_{v}T_{v}\Delta S_{\mathrm{eq},v}>0$.
As a result, the area of rectangle ABOF $\left(T_{p}-T_{c}\right)\Delta S_{\mathrm{eq},c}$
is not equal to the area of rectangle CDEO $\left(T_{h}-T_{p}\right)\Delta S_{\mathrm{eq},h}$,
as shown in Fig. 1(b). 

In the case of a reversible cycle, a heat exchange process occurs
over an infinitely extended duration, i.e., $\tau_{v}\rightarrow\infty$,
leading to $\sum_{v}Q_{v}^{1}$ approaches zero. These quasi-static
isothermal processes ensure that the system remains in thermal equilibrium
with the heat source at all times. It follows from Eq. (\ref{eq:eb})
that $\sum_{v}Q_{v}^{0}$ must be equal to zero. This leads to the
equality between the area of rectangle ABOF and that of rectangle
CDEO, as depicted in Fig. 1(c). At the quasistatic limit, and the
coefficient of performance (COP) of the reversible cycle is then simplified
as follows
\begin{equation}
\psi_{\mathrm{r}}=\frac{Q_{c}}{Q_{h}}=\frac{T_{c}\Delta S_{\mathrm{eq},c}}{T_{h}\Delta S_{\mathrm{eq},h}}=\frac{T_{c}\left(T_{h}-T_{p}\right)}{T_{h}\left(T_{p}-T_{c}\right)}.
\end{equation}

\section{PERFORMANCE OPTIMIZATION IN THE SLOW-DRIVING REGIME}

In the finite-time regime, the COP $\psi$ of the FTQTC can be calculated
as 

\begin{align}
\psi & =\frac{Q_{c}}{Q_{h}}=\frac{T_{c}\left[\Delta S_{\mathrm{eq},c}+\Sigma_{c}/\tau_{c}\right]}{T_{h}\left[\Delta S_{\mathrm{eq},h}+\Sigma_{h}/\tau_{h}\right]},\label{eq:cop}
\end{align}
while the cooling rate is determined by the heat $Q_{c}$ entering
the TLS from reservoir $c$ divided by the total time required for
cooling:

\begin{equation}
R=\frac{Q_{c}}{\tau_{c}+\tau_{h}+\tau_{p}}=\frac{T_{c}\left[\Delta S_{\mathrm{eq},c}+\Sigma_{c}/\tau_{c}\right]}{\tau_{c}+\tau_{h}+\tau_{p}}.\label{eq:R}
\end{equation}
\begin{figure}
\includegraphics[scale=0.3]{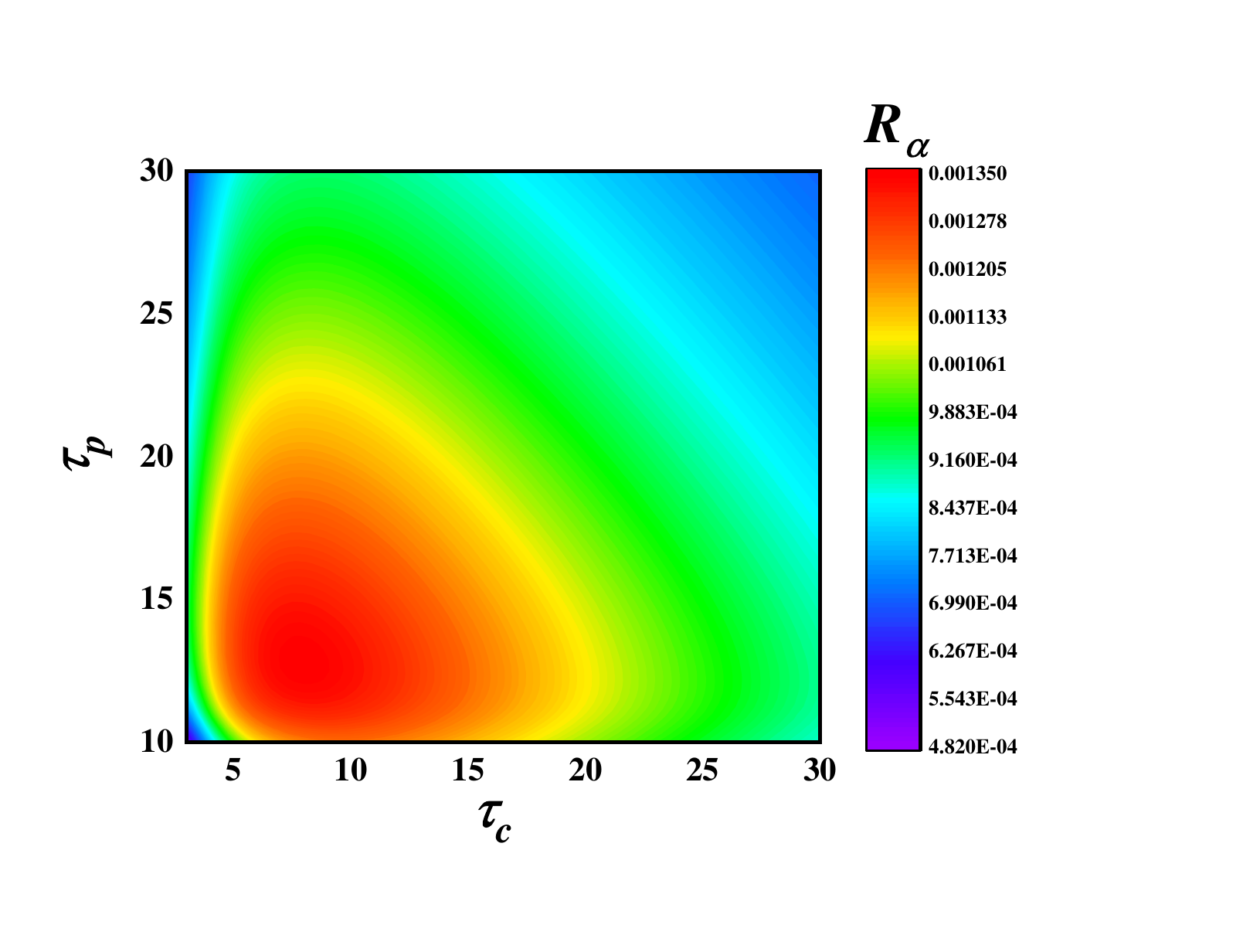}

\caption{Plot of the cooling rate $R_{\alpha}$ with respect to time $\tau_{c}$
and $\tau_{p}$. The other parameters used are the same as those employed
in Fig. 1(c).}
\end{figure}
Here, the duration of the adiabatic process is significantly shorter
than that of the heat exchange process and can be considered negligible.
By utilizing Eqs. ($\ref{eq:cop}$) and (\ref{eq:R}), we can generate
performance characteristic curves for the cooling rate $R$ as a function
of time $\tau_{c}$ and $\tau_{p}$, as illustrated in Fig. 2. It
is clearly seen from Fig. 2 that the cooling rate $R$ is not a monotonic
function of time. The cooling rate can be improved by optimizing the
durations of contact with heat reservoirs. Next, we consider the optimal
configuration of the cycle in which the optimum cooling rate $R$
can be obtained under a given COP $\psi$. By employing the Lagrangian
method \citep{key-Chenjincan1988,key-Wouagfack,key-Chenjincan1996},
we can introduce 
\begin{figure}
\includegraphics[scale=0.3]{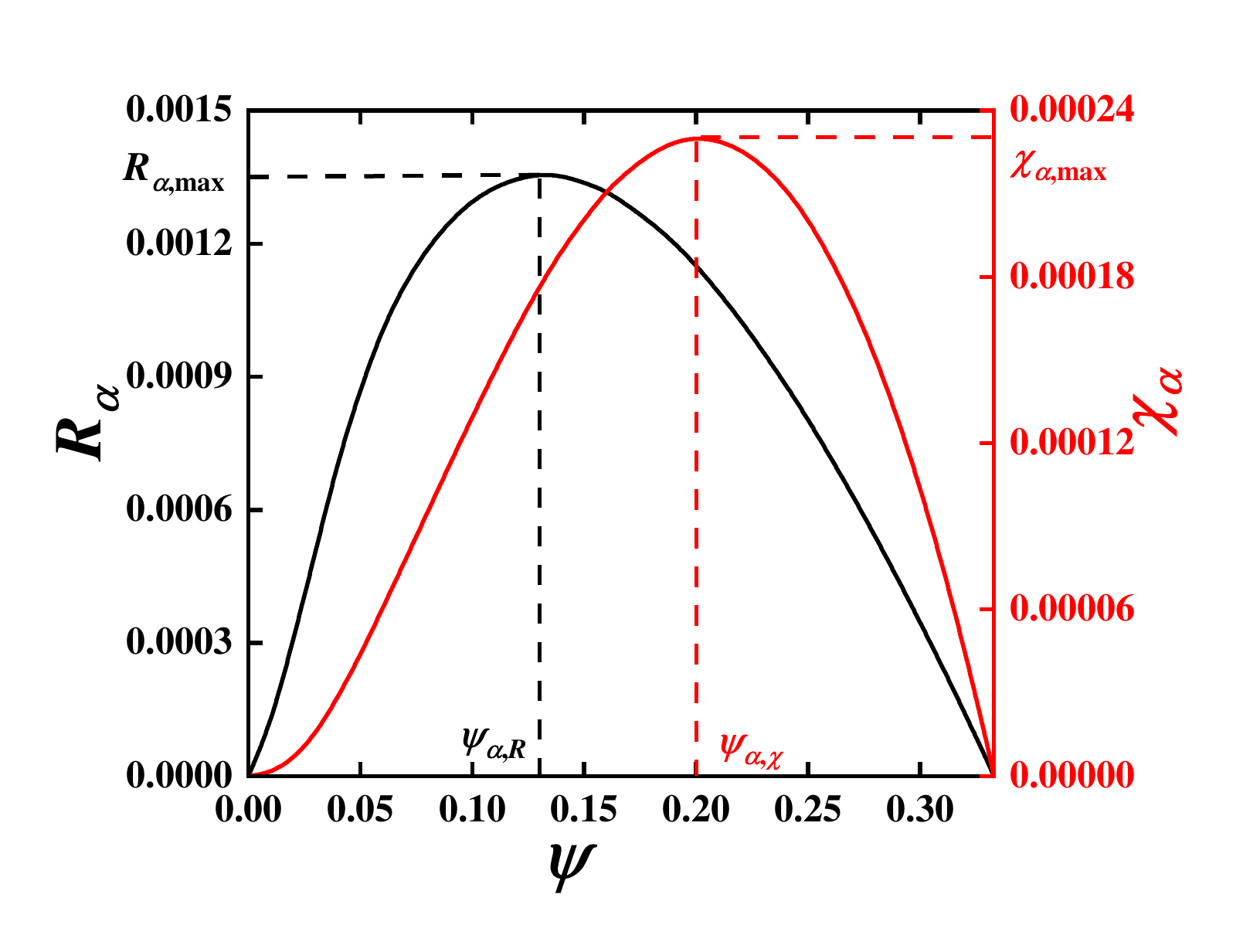}

\caption{The curves of the cooling rate $R_{\alpha}$ (black line) and the
figure of merit $\chi_{\alpha}$(red line) varing with the COP $\psi$.
The values on the left axis correspond to the black lines, whereas
the values on the right axis correspond to the red lines.}
\end{figure}

\begin{align}
L\left(\tau_{c},\tau_{h},\tau_{p}\right) & =R+\lambda_{1}\psi+\lambda_{2}\left(Q_{c}+Q_{h}+Q_{p}\right),\label{eq:L}
\end{align}
where $\lambda_{1}$ and $\lambda_{2}$ are the Lagrange multipliers
associated with the given COP and the law of energy conservation,
respectively. Form the Euler-Lagrange equations, we obtain

\begin{equation}
\frac{\partial L\left(\tau_{c},\tau_{h},\tau_{p}\right)}{\partial\tau_{c}}=0,\label{eq:Lc}
\end{equation}

\begin{equation}
\frac{\partial L\left(\tau_{c},\tau_{h},\tau_{p}\right)}{\partial\tau_{h}}=0,\label{eq:Lh}
\end{equation}
and 
\begin{equation}
\frac{\partial L\left(\tau_{c},\tau_{h},\tau_{p}\right)}{\partial\tau_{p}}=0.\label{eq:Lp}
\end{equation}

According to Eqs.(\ref{eq:Lc}-\ref{eq:Lp}) , the constraint equation
of the times after eliminating the Lagrange multipliers is obtained
by 

\begin{align}
\Delta S_{\mathrm{eq},h}\frac{\tau_{h}^{2}}{\Sigma_{h}}+\Delta S_{\mathrm{eq},p}\frac{\tau_{p}^{2}}{\Sigma_{p}}+\Delta S_{\mathrm{eq},c}\frac{\tau_{c}^{2}}{\Sigma_{c}}+2\left(\tau_{c}+\tau_{h}+\tau_{p}\right) & =0.\label{eq:opt1}
\end{align}
In addition, we have another constraint equation for the time intervals
derived from the energy conservation law as

\begin{equation}
\tau_{h}=-\frac{T_{h}\Sigma_{h}}{T_{p}\left(\Delta S_{\mathrm{eq},p}+\Sigma_{p}/\tau_{p}\right)+T_{c}\left(\Delta S_{\mathrm{eq},c}+\Sigma_{c}/\tau_{c}\right)+T_{h}\Delta S_{\mathrm{eq},h}}.\label{eq:th}
\end{equation}
By combing Eqs. (\ref{eq:opt1}) with (\ref{eq:th}), and considering
$\tau_{c}$ as the independent variable, times $\tau_{h}$ and $\tau_{p}$
for the optimal configuration can be solved.

\section{RESULTS AND DISCUSSION}

By substituting Eqs. (\ref{eq:opt1}) and (\ref{eq:th}) into Eqs.
(\ref{eq:cop}) and (\ref{eq:R}), we can generate the optimal curve
of the cooling rate varying with the COP for given frequency exponent
$\alpha$, as indicated by Fig. 3. The frequency exponent $\alpha$
determines the spectral density $J(\omega_{v}(t))\propto\left[\omega_{v}(t)\right]^{\alpha}$
of the bath. The specific relationship between $\alpha$ and the first
order correction $Q_{v}^{1}$ of heat is elaborated in Appendix C.
As depicted in Fig. 3, the cooling rate $R_{\alpha}$ does not exhibit
a monotonic behavior with respect to the COP. When $\psi=0.13\equiv\psi_{\alpha,R}$,
the cooling rate $R_{\alpha}$ attains its local maximum value $R_{\alpha,\textrm{max}}$.
The quantum refrigerator does not always operate at the state of the
maximum cooling rate. In order to achieve both a higher COP and a
larger cooling rate simultaneously, the refrigerator should be operated
in the region of $\psi\geq\psi_{\alpha,R}$. However, when the refrigerator
is operated in such a region, the cooling rate is a monotonically
decreasing function of the COP. 

It is a worth problem how to choose reasonably the cooling rate and
the COP. The figure of merit $\chi=\psi R$ , originally introduced
by Yan and Chen \citep{key-CJC1990},can be commonly employed as a
target function for optimizing the performance of refrigerators \citep{key-Tomas2012,key-Tu2012,key-PRE107044118(2023)}.
By using Eqs. (\ref{eq:cop}), (\ref{eq:R}), (\ref{eq:opt1}), and
(\ref{eq:th}), we can plot the curves of $\chi_{\alpha}$ as a function
$\psi$ of for a given value of $\alpha$, as represented by the red
solid curve in Fig. 3. It is observed from Fig. 3 that $\chi_{\alpha}$
is not a monotonic function of $\psi$ for a given value of $\alpha$.
When $\psi=0.2\equiv\psi_{\alpha,\chi}$, $\chi_{\alpha}$ attains
its maximum $\chi_{\alpha,\textrm{max}}$. It is seen from Fig.3 that
when $\psi<\psi_{\alpha,R}$, $R_{\alpha}$ decreases with the decrease
of $\psi$. In general, the optimal range of $\psi$ should be determined
by
\begin{figure}
\includegraphics[scale=0.3]{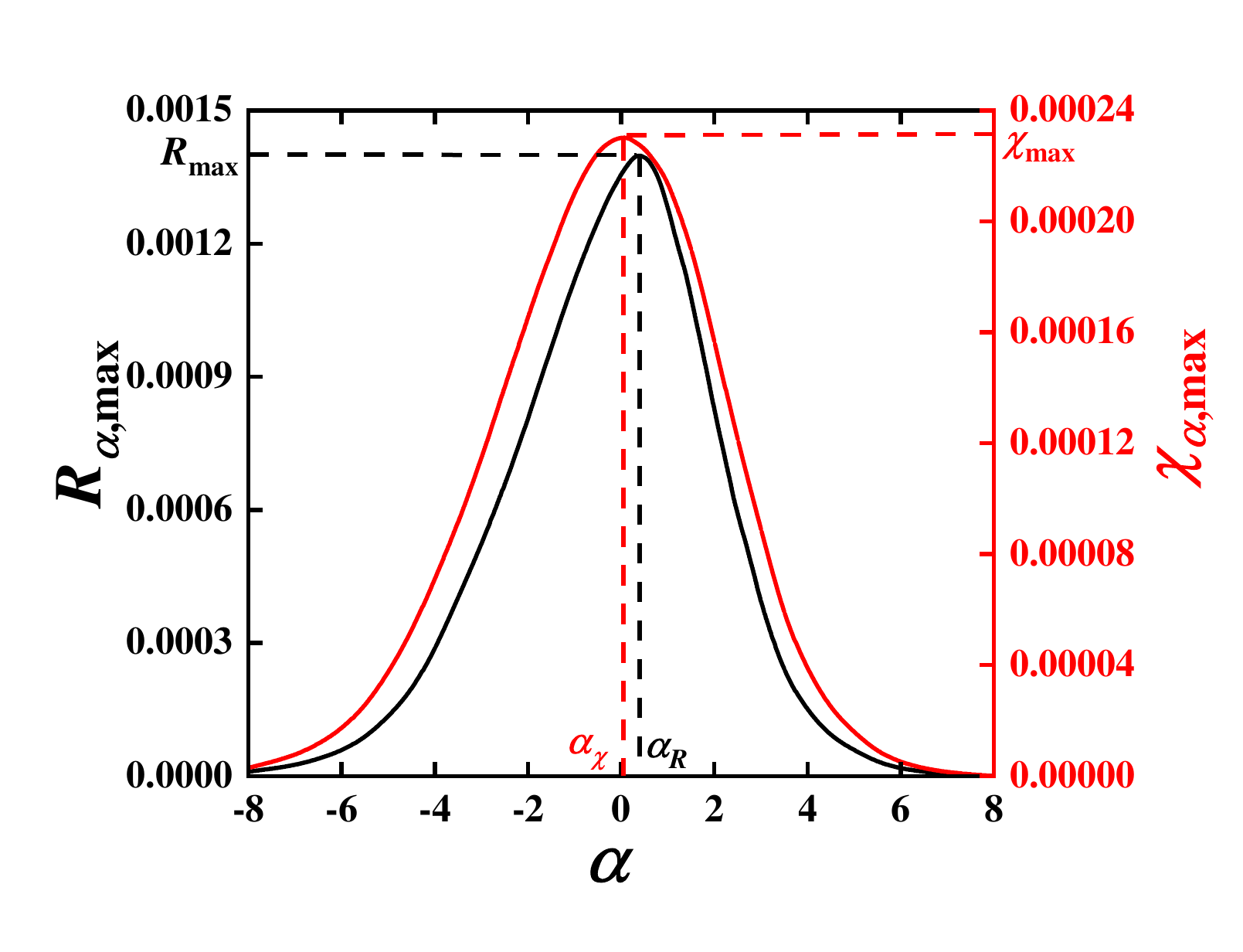}\caption{The cooling rate $R_{\alpha,\textrm{max}}$ and the figure of merit
$\chi_{\alpha,\textrm{max}}$ varying with $\alpha$. The values on
the left axis correspond to the black lines, whereas the values on
the right axis correspond to the red lines.}
\end{figure}
 
\begin{figure}
\includegraphics[scale=0.3]{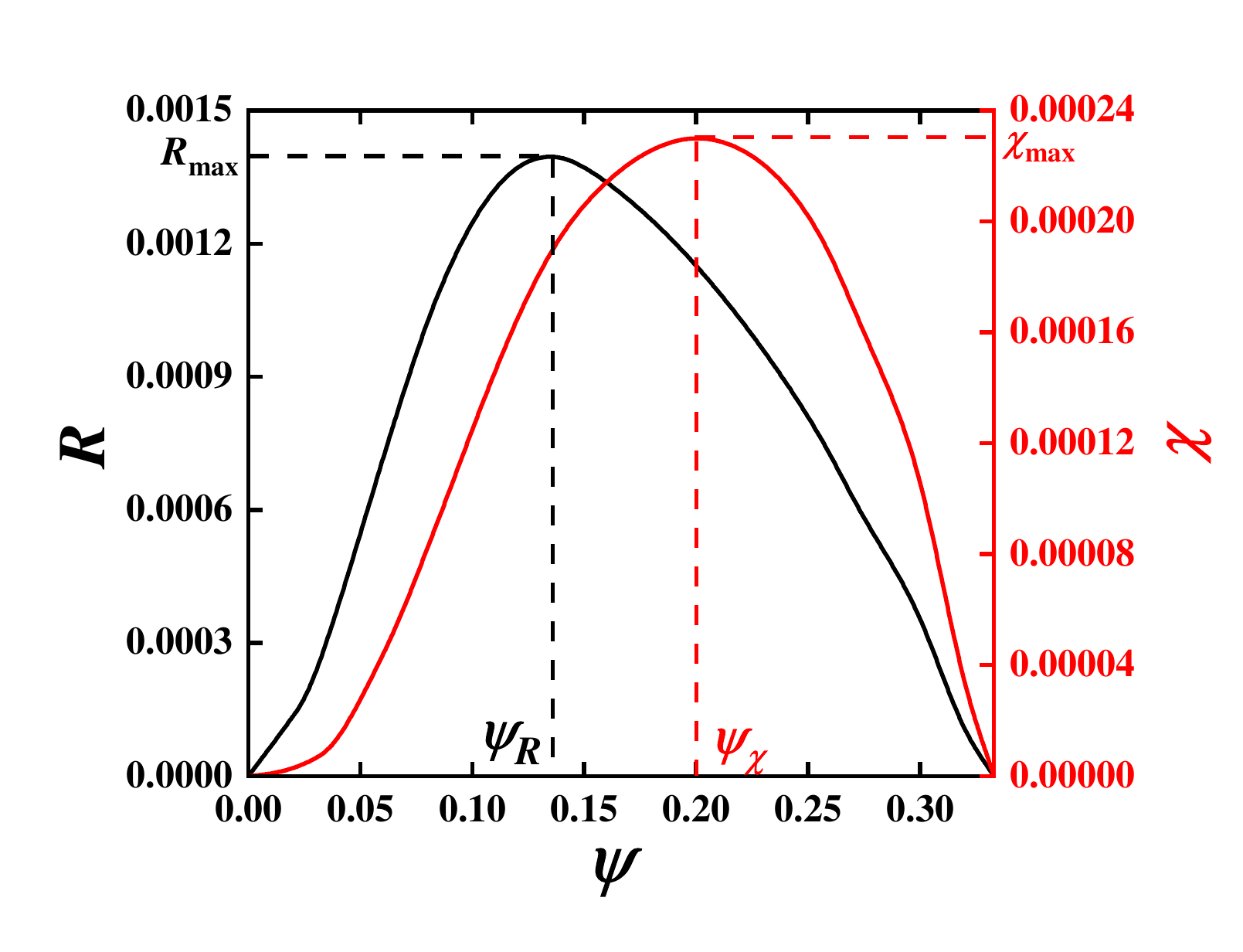}\caption{The optimum characteristic curves of the cooling rate $R$ (black
solid line) and the figure of merit $\chi$ (red solid line) varying
with COP $\psi$. The other parameters used are the same as those
employed in Fig. 1(c).}
\end{figure}

\begin{figure*}
\includegraphics[scale=0.2]{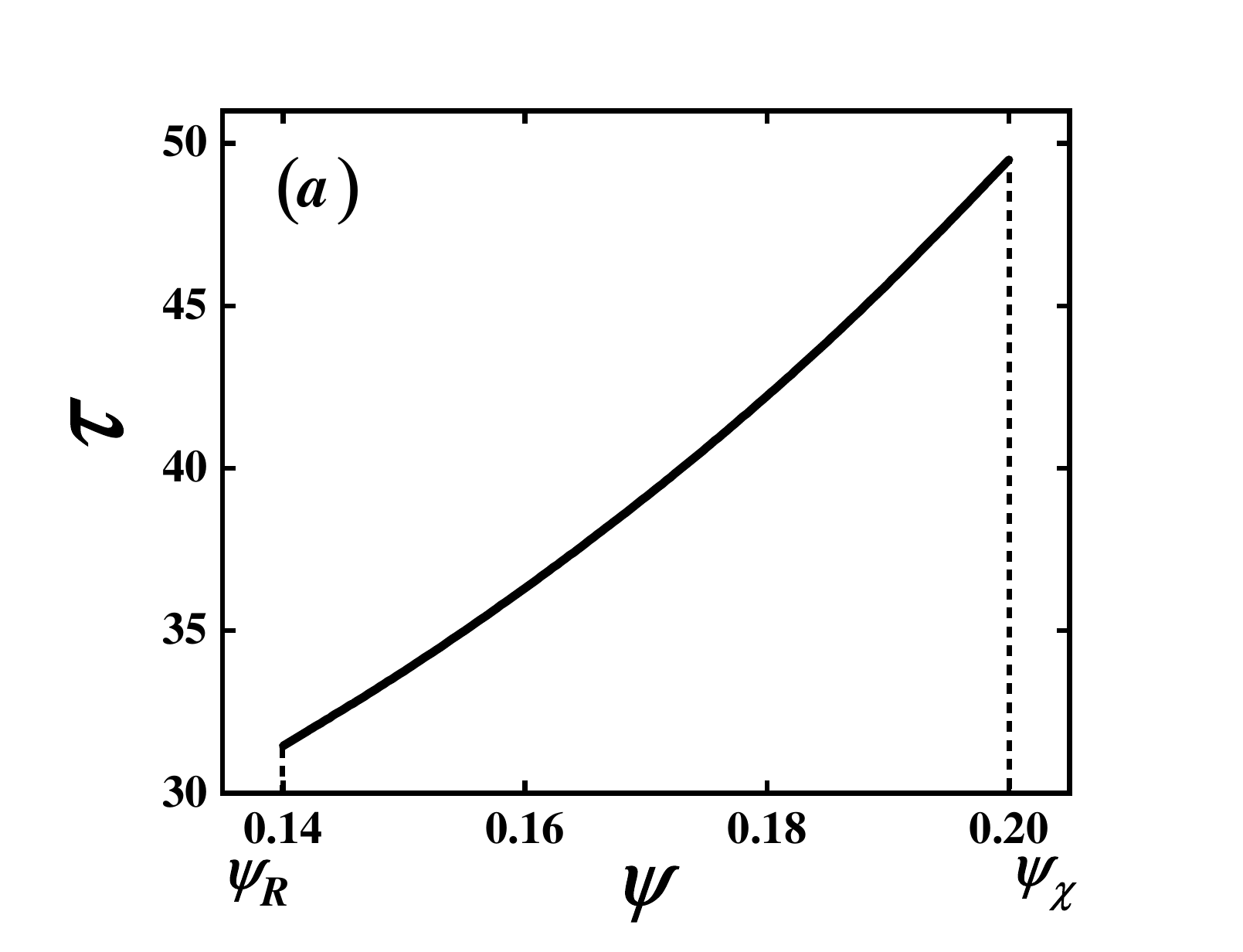}\includegraphics[scale=0.2]{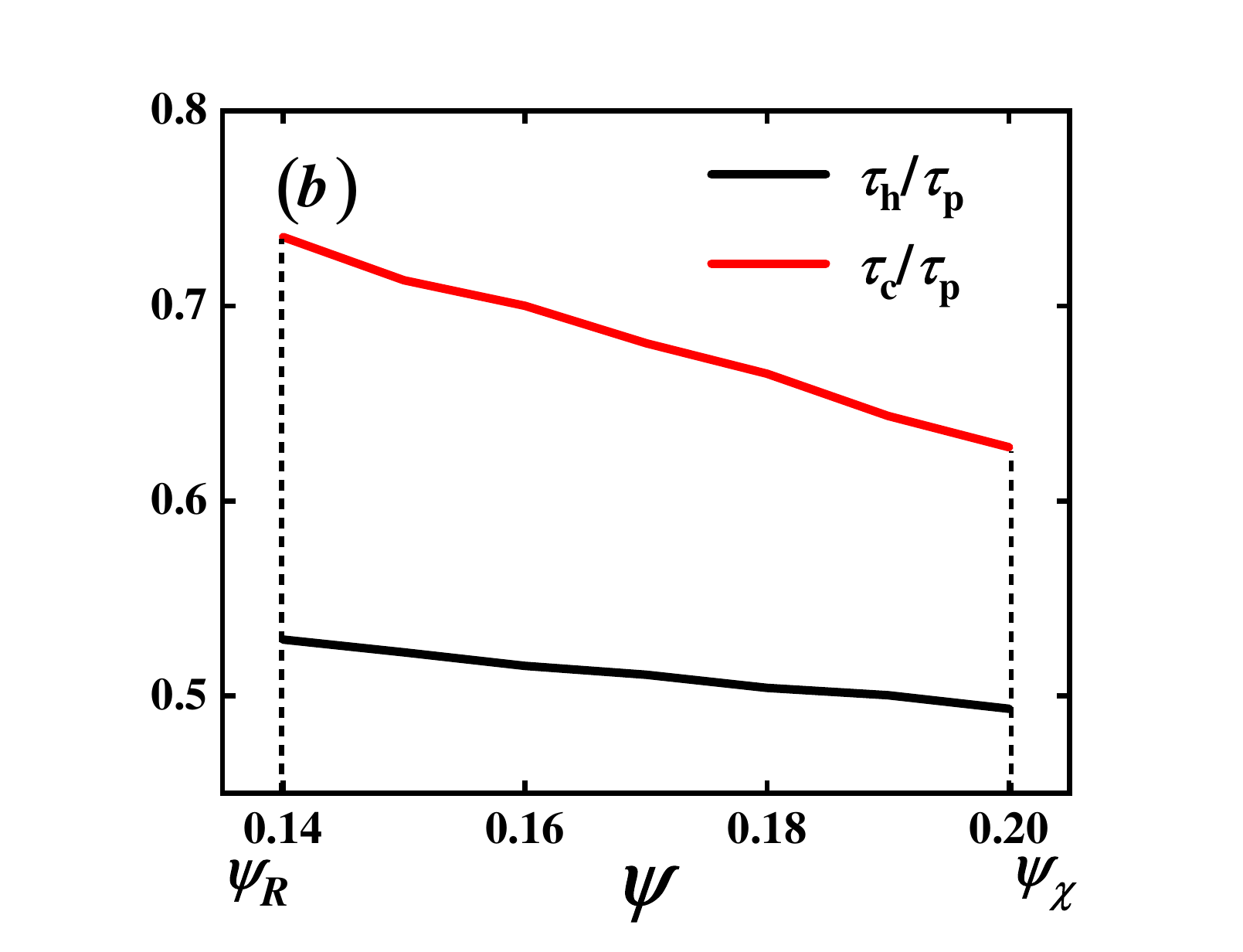}\includegraphics[scale=0.2]{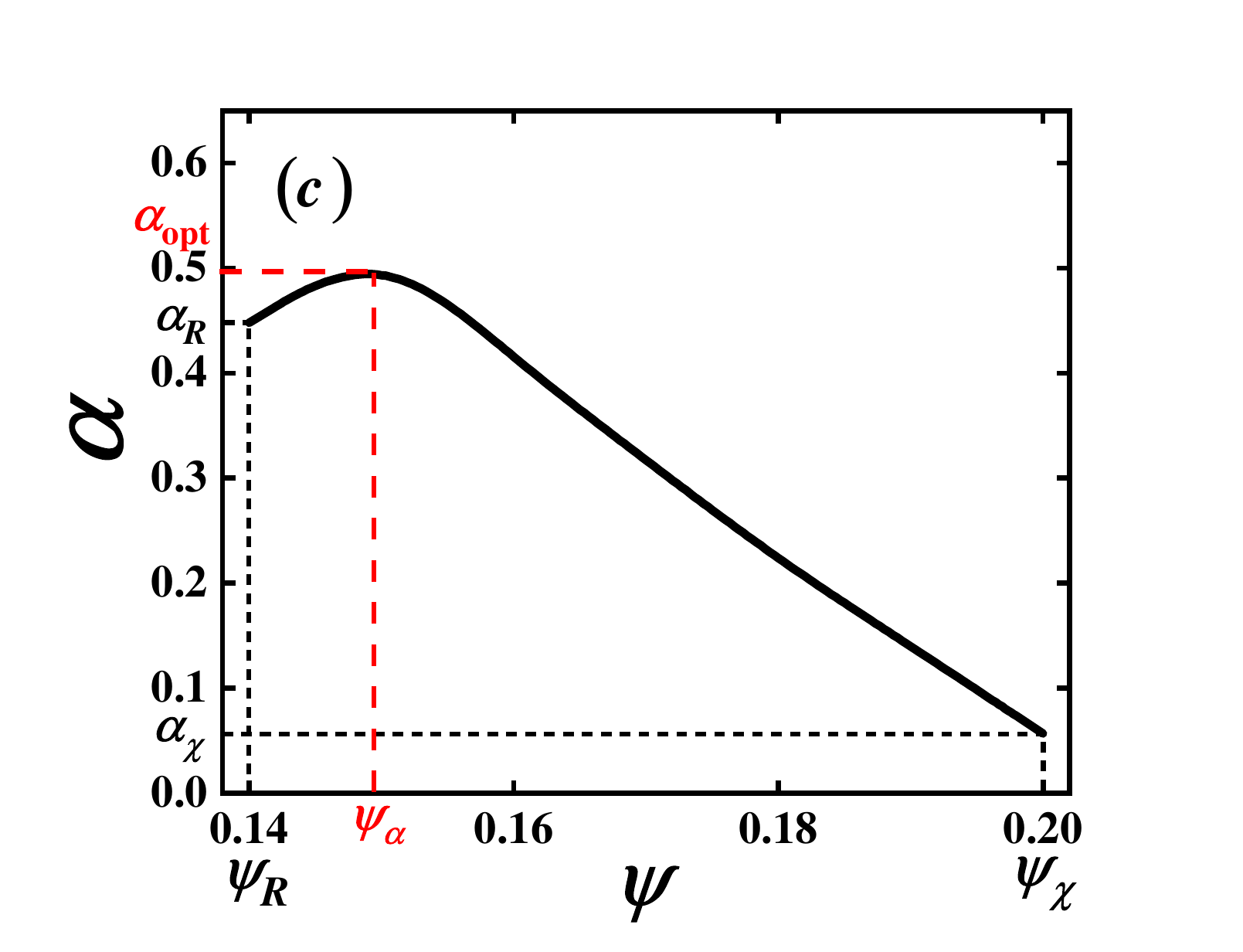}\caption{In the optimal region of the COP, (a) $\tau$ and (b) $\tau_{h}/\tau_{p}$
and $\tau_{c}/\tau_{p}$ and (c) $\alpha$ as functions of $\psi$.
The other parameters used are the same as those employed in Fig. 1(c).}

\end{figure*}

\begin{equation}
\psi_{\alpha,R}\leq\psi\leq\psi_{\alpha,\chi}.
\end{equation}

It can be found through the further analysis that both $R_{\alpha,\textrm{max}}$
and $\chi_{\alpha,\textrm{max}}$ are not monotonic functions of $\alpha$,
as indicated by Fig.4. When $\alpha=0.057\equiv\alpha_{\chi}$, $\chi_{\alpha,\textrm{max}}$
attains its maximum $\chi_{\textrm{max}}$; when $\alpha=0.448\equiv\alpha_{R}$,
$R_{\alpha,\textrm{max}}$ attains its maximum $R_{\textrm{max}}$.
Thus, the optimal range of $\alpha$ should be 

\begin{equation}
\alpha_{\chi}\leq\alpha\leq\alpha_{R}.
\end{equation}

In the region of $\alpha<\alpha_{\chi}$, both $R_{\alpha,\textrm{max}}$
and $\chi_{\alpha,\textrm{max}}$ decrease with the decrease of $\alpha$.
In the region of $\alpha>\alpha_{R}$, both $R_{\alpha,\textrm{max}}$
and $\chi_{\alpha,\textrm{max}}$ decrease with the increase of $\alpha$.
The results obtained here offer a comprehensive framework for efficiently
optimizing the control of a slowly driven FTQTC. 

When $\alpha$ is optimized, we can obtain the $\psi\sim R$ and $\psi\sim\chi$
optimum chararcteric curves, as shown in Fig.5. It is evident from
Fig. 5 that the optimal range of the COP for a FTQTC is given by

\begin{equation}
\psi_{R}\leq\psi\leq\psi_{\chi}.
\end{equation}

This is because both $R$ and $\chi$ decreases with the decrease
of $\psi$ in the region of $\psi<\psi_{R}$, while they decreases
with the increase of $\psi$ in the region of $\psi>\psi_{\chi}$.
In the optimal region of $\psi$, the total cycle time $\tau$ varies
with $\psi$ for two given values $\alpha_{\chi}$ and $\alpha_{R}$
of $\alpha$, as shown in Fig. 6 (a), where two curves almost overlap.
Similarly, we can plot the curves of $\tau_{h}/\tau_{p}$ and $\tau_{c}/\tau_{p}$
varying with $\psi$, as shown in Fig. 6 (b), where two $\psi\sim\tau_{h}/\tau_{p}$
(or $\psi\sim\tau_{c}/\tau_{p}$) curves corresponding to two given
values $\alpha_{\chi}$ and $\alpha_{R}$ of $\alpha$ also almost
overlap. It is observed from Fig.6 (a) and (b) that in the optimal
region of $\psi$, $\tau$ is a monotonically increasing function
of $\psi$, while $\tau_{h}/\tau_{p}$ and $\tau_{c}/\tau_{p}$ are
monotonically decreasing functions of $\psi$. However, $\alpha$
is not a monotonic function of in the optimal region of $\psi$, as
indicated by Fig. 6(c). It is clearly seen from Fig. 6(c) or Eq. (\ref{eq:opt1})
that within three flat baths ($\alpha=0$) are not the optimal selection
for FTQTCs. When a FTQTC is operated within three flat baths, both
the COP and the cooling rate are relatively small. This shows clearly
that when the optimal performance of a FTQTC is researched, it is
not enough to only study the performance of the FTQTCs operated within
three flat baths. 

\section{CONCLUSIONS}

In this work, we presents a finite-time operaton for a quantum tricycle.
The thermodynamic irreversibility during heat exchange processes is
evaluated by analyzing the first-order irreversible corrections of
heat using perturbation theory. The method of Lagrange multipliers
has been employed to determine the optimal performances of the FTQTC.
These configurations determine the cooling rate, the figure of merit,
and the COP at the maximum cooling rate. The results demonstrate that,
for a given spectrum of each bath, the values of the COP corresponding
to the maximum figure of merit and the maximum cooling rate are different.
To simultaneously achieve a relatively large cooling rate and COP,
the performance of the FTQTCs can be improved by operating within
the optimal range of states characterized by the frequency exponent.

\section*{APPENDIX A. SLOW DRIVING OF AN OPEN QUANTUM SYSTEM}

In the case of an open quantum system, where the Hamiltonian $H\left(t\right)$
contains a time-dependent parameter $\omega_{v}\left(t\right)$, and
the system interacts with a reservoir $v$ at an inverse temperature
$\beta_{v}$, we assume that the evolution of the density operator
$\rho(t)$ is governed by the Markovian master equation, 

\begin{equation}
\frac{d}{dt}\rho(t)=\mathscr{\mathscr{\mathcal{L}}_{\mathit{v}}}\left(t\right)\left[\rho(t)\right].\label{eq:Lp-1-1}
\end{equation}
By considering that the Liouvillian operator $\mathscr{\mathscr{\mathcal{L}}_{\mathit{v}}}\left(t\right)$
obeys the quantum detailed balance with respect to $H\left(t\right)$
at all times, the Gibbs state $\rho_{\textrm{eq},v}(t)=\exp[-\beta_{v}H\left(t\right)]/\textrm{Tr}\{\exp[-\beta_{v}H\left(t\right)]\}$
becomes the immediate stationary state of Eq. (\ref{eq:Lp-1-1}),
meaning that 
\begin{equation}
\mathscr{\mathscr{\mathcal{L}}_{\mathit{v}}}\left(t\right)\left[\rho_{\textrm{eq},v}(t)\right]=0.\label{eq:therm}
\end{equation}
As the modulation of frequency $\omega_{v}\left(t\right)$ becomes
infinitely slow, the dynamics of the system follow a quasi-static
isothermal trajectory. During this trajectory, the system gradually
adjusts to the instantaneous equilibrium state near the Gibbs ensemble
$\rho_{\textrm{eq},v}(t)$. Substituting Eq. (\ref{eq:therm}) into
Eq.(\ref{eq:Lp-1-1}), we can derive an alternative evolution equation
for $\rho(t)$ as follows:
\begin{equation}
\frac{d}{dt}\rho(t)=\mathscr{\mathcal{L}}_{\mathit{v}}\left(t\right)\left[\rho(t)-\rho_{\textrm{eq},v}(t)\right].\label{eq:Lp-1}
\end{equation}
As $\rho(t)-\rho_{\textrm{eq},v}(t)$ is a traceless Hermitian operator,
and the action of a superoperator on a traceless subspace is invertible
\citep{key-ScandiPHD}, the solution of Eq. (\ref{eq:Lp-1}) can be
expressed as follows:
\begin{equation}
\rho(t)=\rho_{\textrm{eq},v}(t)+\mathscr{\mathcal{L}}_{\mathit{v}}^{-1}\left(t\right)\left[\frac{d}{dt}\rho_{\textrm{eq},v}(t)\right],\label{eq:pt-16}
\end{equation}
where$\mathscr{\mathcal{L}}_{\mathit{v}}^{-1}\left(t\right)$ is the
Drazin inverse of $\mathscr{\mathcal{L}}_{\mathit{v}}\left(t\right)$
\citep{key-Chenjf,key-Crook,key-Mandal}. Therefore, the solution
for Eq. (\ref{eq:Lp-1}) can be expanded in a series form through
an iterative process. The expansion is given by the following expression:

\begin{equation}
\rho(t)\approx\sum_{n=0}^{\infty}\left(\mathscr{\mathcal{L}}_{\mathit{v}}^{-1}\left(t\right)\frac{d}{dt}\right)^{n}\left[\rho_{\textrm{eq},v}(t)\right].\label{eq:epan}
\end{equation}
Note that in Eq. (\ref{eq:epan}), we have omitted the term $\left(\mathscr{\mathcal{L}}_{v}^{-1}\left(t\right)\frac{d}{dt}\right)^{n+1}\left[\rho(t)\right]$
with $n$ approaching infinity. This omission is justified by assuming
that the derivative of the density operator with respect to time,
when taken an infinite number of times, becomes zero.

We can rewrite Eq. (\ref{eq:epan}) by introducing the dimensionless
time-rescaled parameter $s=t/\tau_{v}$ as follows

\begin{equation}
\tilde{\rho}(s)=\sum_{n=0}^{\infty}\left(\mathscr{\mathcal{\tilde{L}}}_{\mathit{v}}^{-1}\left(s\right)\frac{1}{\tau_{v}}\frac{d}{ds}\right)^{n}[\tilde{\rho}_{\textrm{eq},v}(s)],\label{eq:Lps}
\end{equation}
where $\tilde{\rho}(s)\equiv\rho(\tau s)$ and $\mathscr{\mathcal{\tilde{L}}}_{\mathit{v}}^{-1}\left(s\right)\equiv\mathscr{\mathcal{L}}_{v}^{-1}\left(\tau s\right)$.
Rescaling the time provides the advantage of incorporating the duration
$\tau_{v}$ of the evolution as a straightforward multiplicative factor
in each term of the sum in Eq. (\ref{eq:Lps}). The state of the system
now relies on the time $s$ within the unit interval $s\in[0,1]$,
while the impact of the control protocol on the dynamics is encompassed
within $\mathscr{\mathscr{\mathcal{\tilde{L}}}}_{\mathit{v}}^{-1}\left(s\right)$.
Hence, Eq. (\ref{eq:Lps}) provides a perturbation expansion of the
solution of the density operator in terms of powers of $1/\tau_{v}$.
During the finite-time slow driving process, there exists a lag between
states $\tilde{\rho}(s)$ and $\tilde{\rho}_{\textrm{eq},v}(s)$.
Considering the first-order perturbation, we obtain Eq.(\ref{eq:q1-1})
in the main text.

\section*{APPENDIX B. THERMODYNAMIC QUANTITY OF THE SLOWLY DRIVEN PROCESS}

Building upon the derivation presented in the Appendix A, our objective
is to investigate the impact of deviations from the corresponding
quasistatic trajectory (as described in Eq. (\ref{eq:q1-1})) on the
thermodynamic properties of the process. In order to accomplish this
objective, it is important to observe that the mean energy and von
Neumann entropy of system can be described by the following expressions
as $U(t)=\textrm{\textrm{Tr}}[H(t)\rho(t)]$ and $S(t)=-k_{B}\textrm{\textrm{Tr}}[\rho(t)\log\rho(t)]$,
respectively. In a finite-time process with a duration of $\tau_{v}$,
the change in internal energy is typically divided into two components\citep{key-Alicki2015,key-Alicki1979,key-Liufei}
\begin{equation}
U(\tau_{v})-U(0)=\intop_{0}^{\tau_{v}}dt\textrm{Tr}[H(t)\frac{d}{dt}\rho(t)]+\intop_{0}^{\tau_{v}}dt\textrm{Tr}[\rho(t)\frac{d}{dt}H(t)].\label{eq:U=00003DQ+W}
\end{equation}
Employing Alicki's definition of heat and considering the first-order
perturbation in Eq. (\ref{eq:q1-1}), we can determine the amount
of heat that enters the system from the $\nu$ during the interval
$[0,\tau_{v}]$ as follows:

\begin{align}
Q_{v} & =\intop_{0}^{\tau_{v}}dt\textrm{Tr}[H(t)\frac{d}{dt}\rho(t)]=\intop_{0}^{1}ds\textrm{Tr}[\tilde{H}(s)\frac{d}{ds}\tilde{\rho}(s)],\label{eq:qfenjie}
\end{align}
where the time-rescaled Hamiltonian $\tilde{H}(s)\equiv H(\tau_{v}s)$.
In a quasi-static process that maintains the equilibrium state between
the system and the bath, the zeroth-order approximation in equilibrium
thermodynamics suggests that the system absorbs an amount of heat
given by the expression:
\begin{equation}
Q_{v}^{0}=\intop_{0}^{\tau_{v}}dt\textrm{Tr}[H(t)\frac{d}{dt}\rho_{\textrm{eq},v}(t)]=\intop_{0}^{1}ds\textrm{Tr}[\tilde{H}(s)\frac{d}{ds}\left[\tilde{\rho}_{\textrm{eq},v}(s)\right]].\label{eq:Q0}
\end{equation}
Note that the entropy of the system in the equilibrium state is given
by 
\begin{align}
S_{\textrm{eq\text{,}}v}(t) & =-k_{B}\textrm{Tr}\left[\rho_{\textrm{eq},v}(t)\ln[\rho_{\textrm{eq},v}(t)]\right]\nonumber \\
 & =-k_{B}\textrm{Tr}\left[\tilde{\rho}_{\textrm{eq},v}(s)\ln[\tilde{\rho}_{\textrm{eq},v}(s)]\right].\label{eq:Sv}
\end{align}
Equation (\ref{eq:Q0}) is equivalent to
\begin{equation}
Q_{v}^{0}=\beta_{v}^{-1}\Delta S_{\textrm{eq\text{,}}v},\label{eq:Q0a}
\end{equation}
which is associated with the change in entropy $\Delta S_{\textrm{eq\text{,}}v}=S_{\textrm{eq\text{,}}v}(\tau_{v})-S_{\textrm{eq\text{,}}v}(0)$
along the quasi-static trajectory during the time interval $[0,\tau_{v}]$.
The first-order irreversible correction of heat
\begin{align}
Q_{v}^{1} & =\intop_{0}^{\tau_{v}}dt\textrm{Tr}[H(t)\frac{d}{dt}\left\{ \mathcal{L}_{\mathit{v}}^{-1}(t)\frac{d}{dt}[\rho_{\textrm{eq},v}(t)]\right\} ]\nonumber \\
 & =\tau_{v}^{-1}\intop_{0}^{1}ds\textrm{Tr}\left[\tilde{H}(s)\frac{d}{ds}\left\{ \mathscr{\mathcal{\tilde{L}}}_{\mathit{v}}^{-1}(s)\frac{d}{ds}[\tilde{\rho}_{\textrm{eq},v}(s)]\right\} \right]\nonumber \\
 & =\beta_{v}^{-1}\Sigma_{v}/\tau_{v},\label{eq:Q1}
\end{align}
 where 
\begin{equation}
\Sigma_{v}=\beta_{v}\intop_{0}^{1}ds\textrm{Tr}\left[\tilde{H}(s)\frac{d}{ds}\left\{ \mathscr{\mathcal{\tilde{L}}}_{\mathit{v}}^{-1}(s)\frac{d}{ds}[\tilde{\rho}_{\textrm{eq},v}(s)]\right\} \right]\label{eq:sigma}
\end{equation}
 expresses the increase in dissipation that deviates from the reversible
limit. Based on Appendixes A and B, the dynamics of the TLS under
slow driving will be provided in Appendix C.

\section*{APPENDIX C. THE THERMODYNAMICS OF THE TLS UNDER SLOW DRIVING}

Based on the aforementioned Appendixes, the calculation of the finite-time
correction to the heat absorbed by a TLS is performed. In the weak
coupling regime, the evolution of the system can be described by the
Master Equation as follows:

\begin{align}
\frac{d}{dt}\rho(t) & =\mathscr{\mathscr{\mathcal{L}}_{\mathit{v}}}\left(t\right)\left[\rho(t)\right]=-\frac{i}{\hbar}[H(t),\rho(t)]\nonumber \\
 & \qquad+\gamma_{v}(t)(n_{v}(t)+1)[\sigma_{-}\rho(t)\sigma_{+}-\frac{1}{2}\{\sigma_{+}\sigma_{-},\rho(t)\}]\nonumber \\
 & \qquad+\gamma_{v}(t)n_{v}(t)[\sigma_{+}\rho(t)\sigma_{-}-\frac{1}{2}\{\sigma_{-}\sigma_{+},\rho(t)\}].\label{eq:LS}
\end{align}
Here, the damping rate $\gamma_{v}(t)=\gamma_{0}\left(\omega_{v}(t)\right)^{\alpha}$
relies on the coupling constant $\gamma_{0}$. The frequency exponent
$\alpha$ is determined by the spectral density $J(\omega_{v}(t))\propto\left[\omega_{v}(t)\right]^{\alpha}$
of the bath \citep{key-Cavina,key-Duan}, which is assumed to be identical
for all three baths. The value of $\alpha$ characterizes the dissipation
properties, categorizing it as a flat bath ($\alpha=0$), sub-Ohmic
($0<\alpha<1$), Ohmic ($\alpha=1$), or super-Ohmic ($\alpha>1$)
\citep{key-Cangemi}. The quantity $n_{v}(t)=\left\{ \exp\left[\beta_{v}\hbar\omega_{v}(t)\right]\text{\textminus}1\right\} ^{-1}$
represents the mean number of phonons associated with the bath at
frequency $\omega_{v}(t)$. The notation $\left\{ \star,\ast\right\} $
denotes the anticommutator of the two operators. The operators $\sigma_{+}$
and $\sigma_{-}$ correspond to the raising and lowering operators,
respectively.

The state of the TLS is expressed in terms of the density matrix as
$\rho(t)=\left(\begin{array}{llll}
\rho_{\mathrm{11}} & \rho_{\mathrm{10}} & \rho_{\mathrm{01}} & \rho_{\mathrm{00}}\end{array}\right)^{\mathrm{T}}$, where $\rho_{ij}(t)=\langle i|\rho(t)|j\rangle$ ($i,j=0$ or $1$)
represent the matrix elements of the density matrix. Further, the
superoperator $\mathscr{\mathscr{\mathscr{\mathscr{\mathcal{L}}_{\mathit{v}}}}}(t)$
in the matrix form reads

\begin{widetext}
\begin{equation}
\mathscr{\mathscr{\mathscr{\mathscr{\mathcal{L}}_{\mathit{v}}}}}(t)=\left(\begin{array}{cccc}-\gamma_{v}(t)(n_{v}(t)+1) & 0 & 0 & \gamma_{v}(t)n_{v}(t)\\0 & -\gamma_{v}(t)(n_{v}(t)+\frac{1}{2})-i\omega_{v}(t) & 0 & 0\\0 & 0 & -\gamma_{v}(t)(n_{v}(t)+\frac{1}{2})+i\omega_{v}(t) & 0\\\gamma_{v}(t)(n_{v}(t)+1) & 0 & 0 & -\gamma_{v}(t)n_{v}(t)\end{array}\right).\label{eq:L-1}
\end{equation}
\end{widetext}

At time $t$, the instantaneous equilibrium state can be simplified
as follows:
\begin{equation}
\rho_{\textrm{eq},v}(t)=\left(\begin{array}{cccc}
\frac{n_{v}(t)}{2n_{v}(t)+1} & 0 & 0 & \frac{n_{v}(t)+1}{2n_{v}(t)+1}\end{array}\right)^{\mathrm{T}}.\label{eq:Peq}
\end{equation}
Therefore, the Drazin inverse $\mathcal{L}_{v}^{-1}(t)$ is expressed
as 

\begin{widetext}
\begin{equation}
\mathcal{L}_{v}^{-1}(t)=\left(\begin{array}{cccc}-\frac{n_{v}(t)+1}{\gamma_{v}(t)(2n_{v}(t)+1)^{2}} & 0 & 0 & \frac{n_{v}(t)}{\gamma_{v}(t)(2n_{v}(t)+1)^{2}}\\0 & \frac{1}{-\gamma_{v}(t)(n_{v}(t)+\frac{1}{2})-i\omega_{v}(t)} & 0 & 0\\0 & 0 & \frac{1}{-\gamma_{v}(t)(n_{v}(t)+\frac{1}{2})+i\omega_{v}(t)} & 0\\\frac{n_{v}(t)+1}{\gamma_{v}(t)(2n_{v}(t)+1)^{2}} & 0 & 0 & -\frac{n_{v}(t)}{\gamma_{v}(t)(2n_{v}(t)+1)^{2}}\end{array}\right).\label{eq:L----1}
\end{equation}
\end{widetext}

By employing the time-rescaled parameter $s=t/\tau_{v}$ and utilizing
the zeroth-order approximation in Eq.(\ref{eq:Q0}), the absorbed
heat $Q_{v}^{0}$ from bath $v$ can be estimated in the quasi-static
limit. Furthermore, the first-order correction of the heat $Q_{v}^{1}$
generated by the irreversible finite-time process can be calculated
using Eq.(\ref{eq:Q1}), i.e.,

\begin{equation}
Q_{v}^{1}=\frac{\hbar}{2\tau_{v}}\intop_{0}^{1}\tilde{\omega}_{v}(s)\textrm{Tr}\left\{ \sigma_{z}\frac{d}{ds}\left\{ \mathcal{\tilde{L}}_{v}^{-1}(s)\frac{d}{ds}[\tilde{\rho}_{\textrm{eq},v}(s)]\right\} \right\} ds.\label{eq:Q1a-1}
\end{equation}

\section*{APPENDIX D. THE RELATIONSHIPS BETWEEN THE AMOLITUDE AND THE DISPLACEMENT}

The choice of the parameters, amplitude $\delta_{v}$ and displacement
$\zeta_{v}$, in the alteration of frequency $\omega_{v}\left(t\right)$
during adiabatic operations enables the system to transition from
the equilibrium state associated with one reservoir to the equilibrium
state associated with another reservoir. 
\begin{figure}
\includegraphics[scale=0.3]{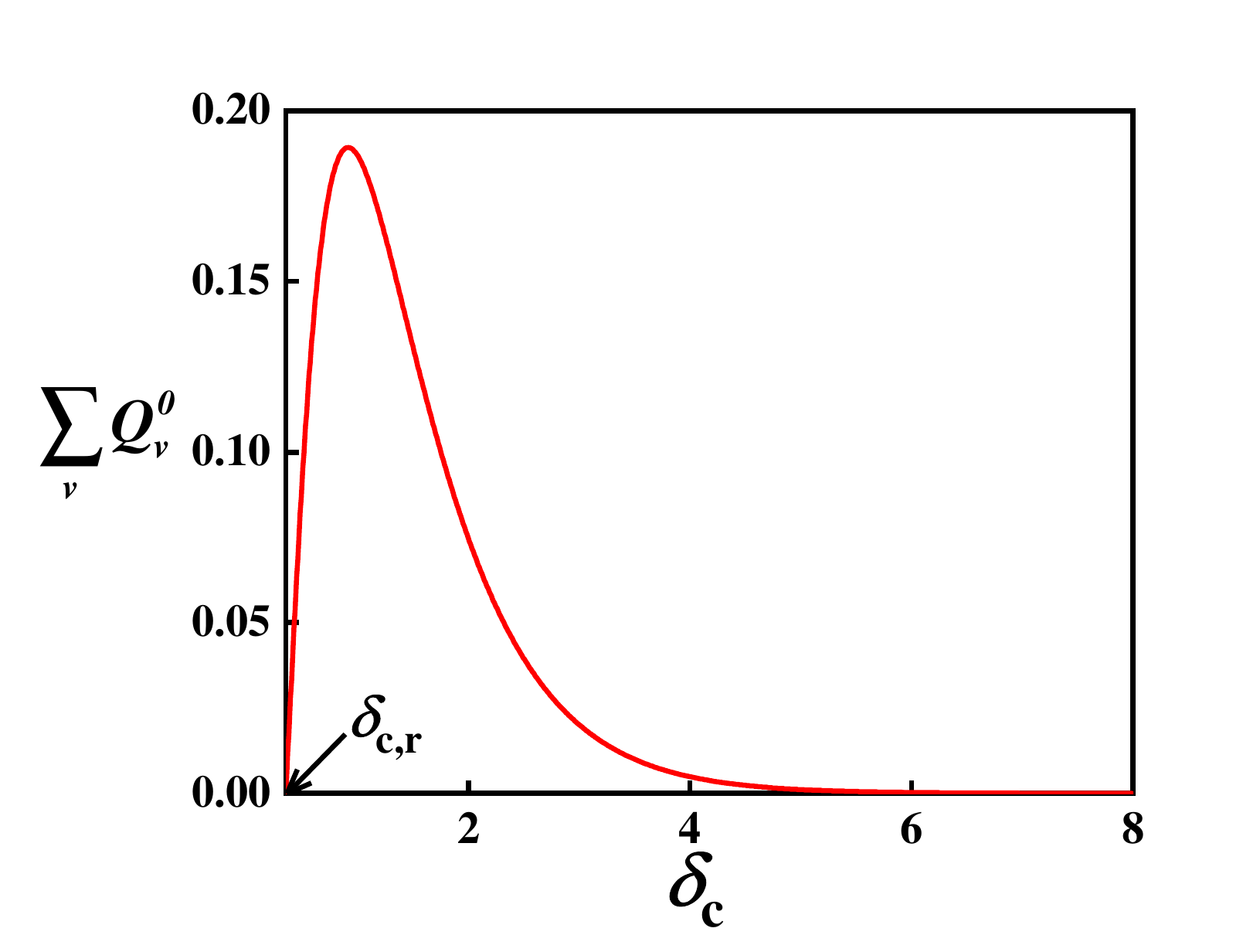}\caption{The curves of the sum of the zeroth order approximation of heat $\sum_{v}Q_{v}^{0}$
varing with the amplitude frequency\textcolor{blue}{{} }$\delta_{c}$.
The remaining parameters utilized are consistent with those employed
in Fig. 1(c).}
\end{figure}
In the context of the quantum tricycle, it is necessary that the frequences
of the TLS at the end of a heat exchange process and at the commencement
of the subsequent heat exchange process are directly proportional
to each other. Specifically, $\omega_{c}\left(\tau_{c}\right)=\left(T_{c}/T_{h}\right)\omega_{h}\left(0\right)$,
$\omega_{h}\left(\tau_{h}\right)=\left(T_{h}/T_{p}\right)\omega_{p}\left(0\right)$,
and $\omega_{p}\left(\tau_{p}\right)=\left(T_{p}/T_{c}\right)\omega_{c}\left(0\right)$.
These requirements lead to the following relationships:

\begin{equation}
\frac{\delta_{c}\left(\zeta_{c}-1\right)}{\delta_{h}\left(\zeta_{h}+1\right)}=\frac{T_{c}}{T_{h}},\label{eq:dc/dh}
\end{equation}

\begin{equation}
\frac{\delta_{h}\left(\zeta_{h}-1\right)}{\delta_{p}\left(\zeta_{p}-1\right)}=\frac{T_{h}}{T_{p}},\label{eq:dh/dp}
\end{equation}
and 

\begin{equation}
\frac{\delta_{p}\left(\zeta_{p}+1\right)}{\delta_{c}\left(\zeta_{c}+1\right)}=\frac{T_{p}}{T_{c}}.\label{eq:dp/dc}
\end{equation}
Employing Eqs. (\ref{eq:dc/dh})-(\ref{eq:dp/dc}), in conjunction
with the provided temperatures $T_{c}$, $T_{h}$, and $T_{p}$, as
well as the displacement values of $\zeta_{c}$ and $\zeta_{h}$,
and treating the amplitude parameter $\delta_{c}$ as the independent
variable, we can deduce the following succinct relationships:

\[
\zeta_{p}=\frac{1+\zeta_{c}\zeta_{h}}{\zeta_{c}+\zeta_{h}},
\]
\[
\delta_{h}=\frac{T_{h}\left(\zeta_{c}-1\right)}{T_{c}\left(1+\zeta_{h}\right)}\delta_{c},
\]
 and 
\[
\delta_{p}=\frac{T_{p}\left(\zeta_{c}+\zeta_{h}\right)}{T_{c}\left(1+\zeta_{h}\right)}\delta_{c}.
\]
In the subsequent discussion, we can investigate an arbitrary quantum
tricycle by varying the value of $\delta_{c}$. 

It is crucial to emphasize that the functioning of the cycle relies
on the condition that the sum value of $\sum_{v}Q_{v}^{0}$ must be
greater than or equal to zero, in accordance with the principles of
energy conversion. The relationship curve in Fig. 7 depicts the variation
of the sum of the zeroth-order approximation of heat $\sum_{v}Q_{v}^{0}$
as a function of the amplitude parameter $\delta_{c}$. It is observed
that the range of the amplitude parameter $\delta_{c}$ for a quantum
tricycle is determined by $\delta_{c}\geq\delta_{c,\textrm{r}}$.
When $\delta_{c}=\delta_{c,\textrm{r}}$, the sum of the zeroth-order
approximation of heat $\sum_{v}Q_{v}^{0}$ is zero, indicating a reversible
tricycle cycle. However, when $\delta_{c}>\delta_{c,\textrm{r}}$,
the sum of the zeroth-order approximation of heat $\sum_{v}Q_{v}^{0}$
is positive, indicating a finite-time irreversible tricycle cycle. 
\begin{acknowledgments}
This work has been supported by the National Natural Science Foundation
(Grants No. 12075197),Natural Science Foundation of Fujian Province
(Grant No. 2023J01006), and the Fundamental Research Fund for the
Central Universities (No. 20720210024). 
\end{acknowledgments}

\end{document}